# Reply to "Comment on 'Unconventional enhancement of ferromagnetic interactions in Cd-doped GdFe2Zn20 single crystals studied by ESR and 57Fe Mössbauer spectroscopies' "

M. Cabrera-Baez [1], J. Munevar [2], R. M. Couto-Mota [1], Y. M. Camejo [1], C. Contreras [3], E. Baggio-Saitovitch [3], M. A. Avila [2] and C. Rettori [2,4]

[1]Departamento de Física, Universidade Federal de Pernambuco, Recife, PE, 50740-560 Brazil

[2]CCNH, Universidade Federal do ABC (UFABC), Santo André, SP, 09210-580 Brazil

[3]Centro Brasileiro de Pesquisas Físicas, Rio de Janeiro, RJ, 2290-180 Brazil

[4]Instituto de Física "Gleb Wataghin," UNICAMP, Campinas, SP, 13083-859 Brazil

(March 25, 2021)

The authors reply to the Comment by P. Canfield et al.

DOI: xxx

The $RT_2Zn_{20}$ family offers an incredible versatility to tune diverse ground states through small modifications of their composition. Different ground states ranging from Pauli paramagnetism[1] ($YCo_2Zn_{20}$), ferromagnetic (FM) ordering[2] ($GdFe_2Zn_{20}$), antiferromagnetism[1] (AFM) ($GdCo_2Zn_{20}$), and heavy fermion properties[3] ($YbFe_2Zn_{20}$) among others, were observed in this family of intermetallic compounds with 86% of Zn[1]. Perturbations of those ground states have the potential to allow the understanding of the involved properties and underlying mechanisms.

We have already explored some members of this intermetallic family from a microscopic point of view using the ESR techniques[4-7], specifically $Y_{1-x}Gd_xCo_2Zn_{20}$, $GdCo_{2-x}Fe_xZn_{20}$, $Yb_{1-x}Gd_xFe_2Zn_{20}$ and $GdFe_2Zn_{20-x}Cd_x$. In our recent publication[7] "*Unconventional enhancement of ferromagnetic interactions in Cd-doped GdFe$_2$Zn$_{20}$ single crystals studied by ESR and $^{57}$Fe Mossbauer spectroscopies*" we have reported an enhancement of the FM transition temperature due to negative chemical pressure, from 86 K to 96 K for $x = 0.0$ and $x = 1.4$, respectively, with an also unexpected, however suspicious reduction of the effective and saturation magnetic moment that was inconsistent with our ESR data[7]. In a comment of our work, by P.C. Canfield et al., they have confirmed our finding about the enhancement of the FM temperature, however, with appreciable differences in the $M(H)$ and $M(T)$ curves for the Cd doped samples. We agree with their analysis of the magnetization data, the saturation of those samples is between $6\mu_B$ and $7\mu_B$ instead of $4\mu_B$ as we have reported. It is indeed likely that we have used an incorrect value of the mass of the Cd doped $GdFe_2Zn_{20}$ samples measured, most likely due to undetected second phase contamination from the non-magnetic Zn flux. Therefore, we agree with the comment by P. C. Canfield et al. about the thorough analysis of the $M(T)$ and $M(H)$ for the Cd doped samples (Figure 2 of their comment).

We also agree with their comment about identical magnetic hyperfine field at zero temperature measured in our $^{57}$Fe Mossbauer experiments for the $x = 0$ and $x = 1.4$ samples, consistent with identical saturated magnetic moments.

We would like to mention that the increase of $T_C$ by the addition of Cd was the challenging motivation for us. This is because the usual RKKY formalism would predict a reduction of $T_C$ as the Gd-Gd distance increases. Since the ESR and Mossbauer experiments are microscopic and local measurements their results do not depend on the mass of the samples, as the magnetization technique does, we would like to emphasize on the importance of our results obtained with these two spectroscopies. That is, the increase of $T_C$ may be associated to the reconstruction of the Fermi surface and/or a new distribution of the d-type of conduction electrons in spite of the negative chemical pressure.

-------------------------------------------------------------------------